# Real-time multimodal image registration with partial intraoperative point-set data

Zachary M C Baum*[1,2], Yipeng Hu[1,2], Dean C Barratt[1,2]

[1]Centre for Medical Image Computing, University College London, London, UK
[2]Wellcome / EPSRC Centre for Interventional and Surgical Sciences, University College London, London, UK
Corresponding Email: zachary.baum.19@ucl.ac.uk

Abstract
We present Free Point Transformer (FPT) – a deep neural network architecture for non-rigid point-set registration. Consisting of two modules, a global feature extraction module and a point transformation module, FPT does not assume explicit constraints based on point vicinity, thereby overcoming a common requirement of previous learning-based point-set registration methods. FPT is designed to accept unordered and unstructured point-sets with a variable number of points and uses a "model-free" approach without heuristic constraints. Training FPT is flexible and involves minimizing an intuitive unsupervised loss function, but supervised, semi-supervised, and partially- or weakly-supervised training are also supported. This flexibility makes FPT amenable to multimodal image registration problems where the ground-truth deformations are difficult or impossible to measure. In this paper, we demonstrate the application of FPT to non-rigid registration of prostate magnetic resonance (MR) imaging and sparsely-sampled transrectal ultrasound (TRUS) images. The registration errors were 4.71 mm and 4.81 mm for complete TRUS imaging and sparsely-sampled TRUS imaging, respectively. The results indicate superior accuracy to the alternative rigid and non-rigid registration algorithms tested and substantially lower computation time. The rapid inference possible with FPT makes it particularly suitable for applications where real-time registration is beneficial.

## 1    Introduction

Multimodal image registration is a fundamental problem in medical imaging research wherein images from different modalities are transformed spatially so that corresponding anatomical structures in each image are aligned. Multimodal image registration, like unimodal registration methods, is historically divided into intensity-based methods and feature-based methods (Hajnal et al., 2001; Viergever et al., 2016). In the literature, these methods are distinguished according to whether the registration seeks to align image features that have been extracted explicitly, for instance, manual or algorithm-based identification of organ boundaries and other anatomical landmarks. In general, points, contours, and surfaces are all commonly used features for registration.

In many multimodal registration applications, such as MR-US or MR-CT alignment, intensity-based registration methods that minimise information-theoretic measures, such as mutual information and normalised mutual information (Gaens et al., 1998; Hermosillo et al., 2002; Lu et al., 2008; Loeckx et al., 2010; Heinrich et al., 2012; Mitra et al., 2012a), or other statistical image similarity metrics (Roche et al., 1998; Liu et al., 2002; Hermosillo et al., 2002; Du et al., 2006; Hahn et al., 2010; De Silva et al., 2013; De Silva et al., 2017) have been widely investigated. Typically, the minimisation of the similarity metric is achieved by an iterative numerical optimization scheme.

Despite the success of intensity-based registration methods, these can perform poorly for input image modalities with very different pixel/voxel intensity characteristics, such as MR and ultrasound. Most saliently, these differences often make it difficult to develop robust intensity-based registration methods that can generalise to different healthcare settings. In such cases, feature-based registration approaches provide a viable alternative for many clinical applications when features, such as organ boundaries, can be defined with minimal user interaction.

Feature-based methods have been widely employed within the field of medical imaging research not only for multimodal image registration methods but for registration in general (Shen and Davatzikos, 2002; Oliveira and Tavares, 2008; Pan et al., 2011; Rasoulian et al., 2012; Wu et al., 2014). This is often due to their simpler and less computationally complex nature with respect to intensity-based methods. Notably, many sparse, surface- point-set matching algorithms require some form of regularization, for example, by using statistical deformable models to permit only physically plausible soft-tissue deformations (Hu et al., 2008; Hu et al., 2011; Hu et al., 2012; Hu et al., 2015; Fedorov et al., 2015; Wang et al., 2016). Additionally, the use of simple data formats, such as point-sets, can provide visually intuitive and easy-to-interpret representations of anatomical structures, which can aid clinical use and be an effective basis for clinical user interaction, such as manual refinement (Mani and Arivazhagan, 2013), providing feedback on registration uncertainty and quality (Hu et al., 2016).

Feature extraction has seen rapid advances in recent years given the development of automatic, well-validated, learning-based medical image segmentation methods. Such methods can yield real-time delineation of anatomical surfaces (Ronneberger et al., 2015). These surfaces may be sampled into point-sets for surface matching, for example, using classical point-set registration algorithms, such as the Iterative Closest Point (ICP) (Besl and McKay, 1992). More contemporary alternatives, such as Coherent Point Drift (CPD) (Myronenko and Song, 2007) and Thin-Plate Spline Robust Point Matching (Chui and Rangarajan, 2003) provide a solution for non-rigid registration. Gaussian mixture models (GMM) have been used to compute registrations using probabilistic point correspondences (Jian and Vemuri, 2010).

Recently, several deep-learning-based medical image registration methods have been described (Hu et al., 2018; Fu et al., 2021; Hu et al., 2021). These can also be divided into intensity- and feature-based approaches, with the distinction apparent from whether the input data are image features (e.g. (Hansen et al., 2019; Baum et al. 2020; Fu et al., 2021)), or images (e.g. (Yan et al., 2018; Hu et al., 2018; Haskins et al., 2019)). Convolutional artificial neural networks have also been used to perform medical image registration by learning similarity metrics directly from the images (Yan et al., 2018; Haskins et al., 2019), through image synthesis methods that convert the appearance of one or both input modalities such that they closely resemble the other before registration (Onofrey et al., 2016; Cao et al., 2016; Xu et al., 2020), and through reinforcement learning (Ma et al., 2017; Sun et al., 2018; Hu et al., 2021). Image segmentation data may be used to learn non-rigid statistical deformation models which may, at inference, be used to guide a non-rigid surface registration (Onofrey et al., 2017). Segmentations have been used to determine the correspondence between different imaging modalities in a weakly-supervised framework, with the advantage that the input images are only required at inference (Hu et al., 2018).

Although widely used, classical iterative feature-based registration methods, such as ICP, are not well-suited for applications requiring real-time registration since they are computationally intensive when processing large point/surface datasets. In contrast, the computationally efficient nature of deep neural networks has motivated their application to real-time registration (Aoki et al., 2019; Liu et al., 2019; Wang and Solomon, 2019a; Wang and Solomon, 2019b; Kurobe et al., 2020). Several such methods (for example, (Aoki et al., 2019; Liu et al., 2019)) have exploited PointNet (Qi et al., 2017), a deep learning framework for the classification and segmentation of point-sets. One such method, developed by Aoki et al. (2019), combined PointNet with the Lucas and Kanade algorithm to create an iterative, rigid point-set registration algorithm. Other works have applied PointNet as a means to learn hierarchical features to their method for 3D scene flow (Liu et al. 2019). Without using PointNet, Wang and Solomon (2019a, 2019b) presented methods that provide iterative, self-supervised, rigid registration of partial point-sets with Partial Registration Networks (PRNet), and rigid registration as a pre-cursor to ICP with Deep Closest Point (DCP). Kurobe et al. (2020) developed an approach that regressed correspondence between point-sets by using local and global features to compute the singular value decomposition for rigid registration. Interestingly, these methods all rely on constrained transformation models or loss functions to model noise, outliers, and missing data. Additionally, while the above mentioned methods have been reported for rigid or affine registration of point-sets, some non-rigid registration methods have also been proposed and applied exclusively to medical image registration for the analysis of lung motion (Hansen et al., 2019) and multimodal image registration (Fu et al., 2021).

In this work, we describe a deep neural network architecture for non-rigid point-set registration, which we call Free Point Transformer (FPT). The network consists of two parts: a global feature extraction module and a point transformation module. Importantly, FPT is not limited by the inherently unordered structure of point-sets and predicts a non-rigid transformation that aligns them. To investigate the application of FPT for the registration of partial volumetric point-sets comprising points extracted from MRI and transrectal ultrasound (TRUS) images. This exemplar application illustrates a common situation with applications in which real-time, interventional imaging, such as TRUS, is used where partial (and potentially noisy) point data is available, in this case, because of the use of 2D US imaging. This work aimed to compare the accuracy and speed of FPT-based point-set registration with alternative methods.

Initial results were presented in our preliminary work (Baum et al., 2020). The work presented here expands on this preliminary work in several ways, and we outline our contributions as follows:

- We provide a detailed description of our methodology for unsupervised point-set registration; a method that accepts unordered and unstructured point-sets with a variable number of points.
- We introduce our "model-free" approach which allows non-rigid registration using data-driven learning without known correspondence or heuristic constraints.
- We introduce and describe the implementation and training strategies of the two modules contained within our method; which transforms points that are independent of those which define the registration and, as a result, enable various types of practically useful applications.
- We present rigorous analysis validation experiments which compare our registration methodology, different learning-based methods, and classical iterative point-set registration methods for different clinical scenarios for the prostate MR-TRUS image registration application, including a set of partial-data experiments by varying levels of data availability.

## 2 Methods

### 2.1 Free Point Transformer

Given a pair of source and target point-sets, $\{\mathbf{p}_s \mid s = 1, \dots, N_s\}$ and $\{\mathbf{p}_t \mid t = 1, \dots, N_t\}$, respectively, where $\mathbf{p}_s$ and $\mathbf{p}_t$ are $D$-dimensional vectors denoting individual point spatial coordinates in x-, y-, and z directions (here, $D = 3$). The FPT framework aims to learn a model for inferring a transformation function $\mathcal{T}^{\{\mathbf{p}_s\} \to \{\mathbf{p}_t\}}$, between a pair of point-sets, such that it will map any new point, represented by the vector $\widetilde{\mathbf{p}}_s$ in source coordinate space, to the target coordinates $\widetilde{\mathbf{p}}'_s$ as follows:

$$\widetilde{\mathbf{p}}'_s = \mathcal{T}^{\{\mathbf{p}_s\} \to \{\mathbf{p}_t\}}(\widetilde{\mathbf{p}}_s) \text{ (eq.1)}$$

$\widetilde{\mathbf{p}}_s$ is usually sampled from the source point-set domain, but not necessarily an element of $\{\mathbf{p}_s\}$.

FPT models such a spatial point transformation using a parametric neural network $\mathcal{T}_{\boldsymbol{\theta}}^{\{\mathbf{p}_s\}\rightarrow\{\mathbf{p}_t\}}(\widetilde{\mathbf{p}}_s)$, with network parameters $\boldsymbol{\theta}$, together with an end-to-end network training approach. The FPT network contains two modules: a *global feature extraction* module and a *point transformation* module. The global feature extraction module converts point-sets into a feature vector, whereas, the point transformation module predicts a displacement vector for the given input point $\widetilde{\mathbf{p}}_s$ using the feature vector. A detailed illustration of the two modules and the network training scheme is shown in Figure 1. In the following sections, we provide details on how these two modules are constructed and simultaneously trained using a training set consisting of examples of different point-set pairs.

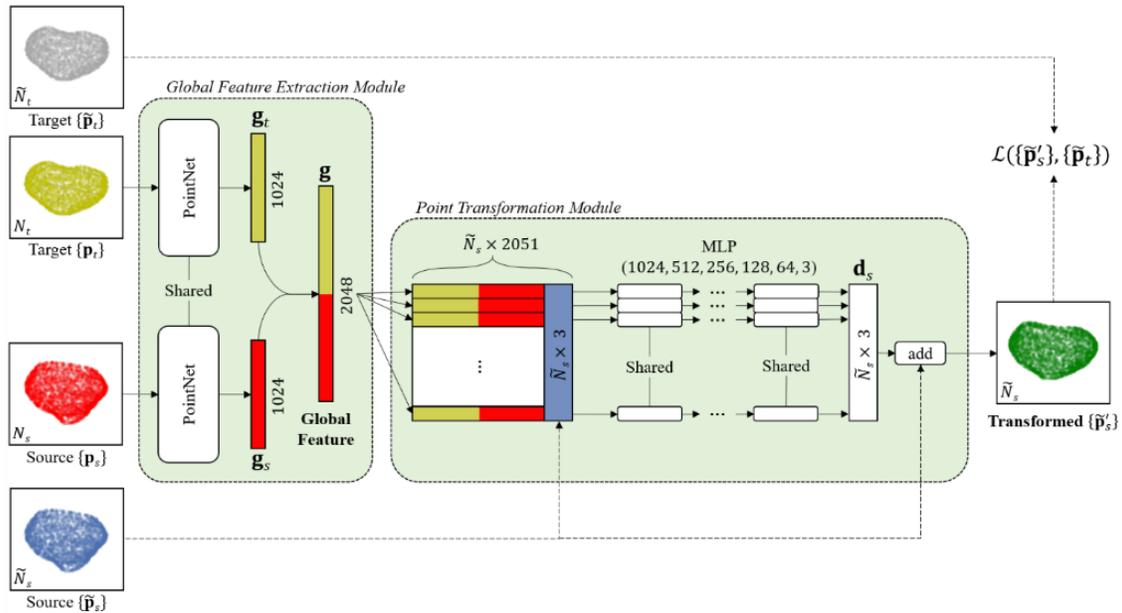

Figure 1: Schematic representation of the FPT network design for non-rigid point-set registration. The global feature extraction module takes a target and source point-set and applies shared input and feature transformations to both, creating a global feature vector. The point transformation module serves as a per-point transformation of the source point-set by determining the displacement to be added in order to obtain the transformed point-set. MLP stands for multi-layer perceptron.

Point-sets have important attributes which we have exploited in the design of the FPT, and which deliver several advantages for registration purposes: First, FPT accepts unordered and unstructured point-sets with a variable number of points. This requires the global feature extraction module to learn a representation, which determines a permutation, rotation, and cardinality invariant feature extraction step. The global feature extraction module adapts the previously proposed PointNet architecture (Qi et al., 2017) to register a pair of point-sets. Second, the FPT has two separate functions: i) predicting a transformation from the network input, registration of the two input point-sets; ii) predicting displacements for individual given points. These two functions are implemented with the global feature extraction module and the point transformation module, which are trained together but may be used for independent point-sets – i.e., those to register $\{\mathbf{p}_s\}$ and $\{\mathbf{p}_t\}$ – and those to be transformed $\{\widetilde{\mathbf{p}}_s\}$. This flexibility is important as it allows the network inputs to be different from the point-sets used for computing the loss, which may only be available during training. Third, the point transformation module in the proposed FPT is defined without an explicit or parameterized registration method, permitting a "model-free" approach. This leads to non-rigid registration using a data-driven learning approach that prevents any collapse or folding that arguably may not be reflective of the data. FPT is trained without heuristic constraints, such as deformation smoothness or hand-engineered noise models. Training in this manner may ultimately be beneficial, as the restrictions imposed by such constraints or models enforce deformation that may over-simplify the complex soft tissue deformation and observable inter-structure motion. As a result of these attributes and considerations, FPT is versatile and permits generalization to partial data while learning from complete data. The FPT supports different types of learning supervision, including fully-supervised, semi-supervised, and partially- or weakly-supervised training (see also a brief discussion in Section 2.2). However, in this work, we focus on training the FPT using an unsupervised learning approach, which allows it to learn from raw point-set data without the need for ground-truth transformations. As we demonstrate for our exemplar use-case, this allows an end-to-end process to be achieved, which includes data acquisition followed by registration in real-time, an ability that is very important for many time-critical medical applications of image registration.

### 2.1.1 Global Feature Extraction Module

PointNet (Qi et al., 2017) was originally designed to convert point-sets into permutation and rotation invariant feature vectors for classification and segmentation tasks. From the original PointNet architecture (Qi et al., 2017), we utilized the input and feature transformation and global information aggregation components to create high-dimensional feature vectors. Unlike the original PointNet architecture, which learns a $3 \times 3$ transformation matrix and subsequently multiplies this learned transform by the coordinates of the input points (Qi et al., 2017), FPT's global feature extraction module learns a $4 \times 4$ transformation matrix to better allow for the representation of 3D translation in homogeneous coordinates, in addition to any rotation, scaling, shearing or reflections which may be represented in the original $3 \times 3$ transformation matrix. As in PointNet, this $4 \times 4$ transformation matrix is then used to transform the coordinates of the input points. This modification resulted from initial experimental results wherein a lower translational error was observed when the adapted PointNet was given the ability to encode translational differences more easily between point-sets in its feature representations. Additionally, batch normalization layers were removed from the PointNet to prevent the normalization of translational differences between source and target point-sets. In FPT, the above modifications create a single PointNet shared between the input point-sets $\{\mathbf{p}_s\}$ and $\{\mathbf{p}_t\}$, as illustrated in Figure 1. The module, in turn, generates feature vectors $\mathbf{g}_s$ and $\mathbf{g}_t$ with pre-defined lengths, from the source and target $\{\mathbf{p}_s\}$ and $\{\mathbf{p}_t\}$, respectively.

$$\mathbf{g} = f_{\boldsymbol{\theta}_{feat}}^{\{\mathbf{p}_s\} \to \{\mathbf{p}_t\}} \text{ (eq.2)}$$

where $\mathbf{g} = [\mathbf{g}_s^T, \mathbf{g}_t^T]^T$ is the concatenated $K$-dimensional global feature vector and $f_{\boldsymbol{\theta}_{feat}}^{\{\mathbf{p}_s\} \to \{\mathbf{p}_t\}}$ denotes the modified PointNet that represents a set-order-sensitive feature extraction function, i.e. $f_{\boldsymbol{\theta}_{feat}}^{\{\mathbf{p}_s\} \to \{\mathbf{p}_t\}} \neq f_{\boldsymbol{\theta}_{feat}}^{\{\mathbf{p}_t\} \to \{\mathbf{p}_s\}}$, which is invariant to the point-order in each set. $\boldsymbol{\theta}_{feat}$ are the network parameters in the global feature extraction module.

### 2.1.2 Point Transformation Module

The point transformation module serves as a per-point transformation model $f_\mathbf{g}$ that predicts the displacement vector that transforms a point $\widetilde{\mathbf{p}}_s$ to $\widetilde{\mathbf{p}}'_s$, conditioned on the computed global feature vector $\mathbf{g}$ (eq.2):

$$\widetilde{\mathbf{p}}'_s = f_{\boldsymbol{\theta}_{trans}}(\widetilde{\mathbf{p}}_s | \mathbf{g}) \text{ (eq.3)}$$

In this work, we use a multi-layer perceptron (MLP) network to model this transformation with network parameters $\boldsymbol{\theta}_{trans}$. Without loss of generality, the hidden units at $l^{th}$ layer in a $L$-layer MLP, $\mathbf{x}^{(l)} = \left[x_j^{(l)}\right]^T, j = 1, \ldots, J^{(l)}$, representing the output feature vector with $J^{(l)}$ ($l = 1, \ldots, L$) elements can be given in a recursive form:

$$x_j^{(l)} = a^{(l)} \left( \sum_{j=1}^{J^{(l-1)}} w_j^{(l)} x_j^{(l-1)} + w_0^{(l)} \right) \text{ (eq.4)}$$

where $a^{(l)}$ is the element-wise activation function (rectified linear units are used in this work); and $w_j^{(l)}$ ($j = 1, \ldots, J^{(l-1)}$) are the weights for each of the $J^{(l-1)}$ elements in the input feature vector $\mathbf{x}^{(l-1)} = \left[ x_j^{(l-1)} \right]^T$ ($j = 1, \ldots, J^{(l-1)}$) from the previous layer. Together with the scalar bias weight $w_0^{(l)}$, the point transformation module parameters are $\boldsymbol{\theta}_{trans} = \left[ \left[ w_j^{(l)} \right]^T_{j=0,1,\ldots,J^{(l-1)}} \right]^T_{l=1,\ldots,L}$.

The point transformation module $f_{\boldsymbol{\theta}_{trans}}$ is specified by the module input and output, the point-concatenated global feature vector $\mathbf{x}^{(0)} = [\mathbf{g}^T, \widetilde{\mathbf{p}}_s^T]^T$ and the displacement vector $\mathbf{d}_i = \mathbf{x}^{(L)}$, respectively; therefore, $J^{(0)} = K + 3$ and $J^{(L)} = 3$. The transformed point can be computed by $\widetilde{\mathbf{p}}'_s = \widetilde{\mathbf{p}}_s + \mathbf{d}_i$. Predicting the displacement $\mathbf{d}_i$, instead of the transformed points $\widetilde{\mathbf{p}}'_s$ directly, which we found empirically simplified the initialisation of the model training. It is important to note that the transformation model parameterised by the above-described MLP does not have constraints on the transformation smoothness, which are commonly imposed with assumptions such as coherence between adjacent points, giving a less constrained transformation.

The use of MLP parameterisation also facilitates an efficient one-dimensional (1D) convolution implementation for multiple feature vectors during the FPT network training. For each network input point-set pair, $\{\mathbf{p}_s\}$ and $\{\mathbf{p}_t\}$, the global feature extraction module computes a global feature vector $\mathbf{g}$ (Eq. 2). In the general case, the point transformation module aims to transform a point-set $\{\widetilde{\mathbf{p}}_s\}, s = 1, \ldots, \widetilde{N}_s$, using Eq. 3, conditioned on the same global feature vector $\mathbf{g}$. Assume a row-wise concatenated "feature matrix" $\mathbf{M}^{(l)}$, $l=1,\ldots L$, at $l^{th}$ layer, such that:

$$\mathbf{M}^{(l)} = \begin{bmatrix} \left( \mathbf{x}_{s=1}^{(l)} \right)^T \\ \left( \mathbf{x}_{s=2}^{(l)} \right)^T \\ \vdots \\ \left( \mathbf{x}_{s=\widetilde{N}_s}^{(l)} \right)^T \end{bmatrix}, \text{ where } \mathbf{M}^{(0)} = \begin{bmatrix} [\mathbf{g}^T, \widetilde{\mathbf{p}}_{s=1}^T] \\ [\mathbf{g}^T, \widetilde{\mathbf{p}}_{s=2}^T] \\ \vdots \\ [\mathbf{g}^T, \widetilde{\mathbf{p}}_{s=\widetilde{N}_s}^T] \end{bmatrix} \text{ and } \mathbf{M}^{(L)} = \begin{bmatrix} \mathbf{d}_{i=1}^T \\ \mathbf{d}_{i=2}^T \\ \vdots \\ \mathbf{d}_{i=\widetilde{N}_s}^T \end{bmatrix}$$

Now, for computing the output feature vector at the $l^{th}$ layer, substituting the network weight $w_j^{(l)}$ in Eq.4 with a scalar weight $k_j^{1,(l)}$, the $j^{th}$ of the $J^{(l-1)}$ 1D convolution kernels for each of the $J^{(l)}$ elements. The $J^{(l-1)} \times J^{(l)}$ kernels are convolved over all $\widetilde{N}_s$ elements in the column space of the feature matrices $\mathbf{M}^{(l)}$ because the MLP weights are shared between all the input row vectors $[\mathbf{g}^T, \widetilde{\mathbf{p}}_s^T]$ in the feature matrices. The rows representing different feature-vector-concatenated points in $\{\widetilde{\mathbf{p}}_s\}$ remain independently multiplied by the 1D kernel.

*2.2 FPT Network Training*

The FPT network described here were trained to optimise the network parameters $\boldsymbol{\theta} = \left[ \boldsymbol{\theta}_{feat}^T, \boldsymbol{\theta}_{trans}^T \right]^T$ by minimising the distance between the transformed source point-set $\{\widetilde{\mathbf{p}}'_s\}$ and a given target point-set $\{\widetilde{\mathbf{p}}_t\}, t = 1, \ldots, \widetilde{N}_t$. The specific form of the function $\mathcal{L}(\{\widetilde{\mathbf{p}}'_s\}, \{\widetilde{\mathbf{p}}_t\} | \boldsymbol{\theta})$ serves as the training loss and is described in Section 2.2.1, while the goal of the network training is:

$$\widehat{\boldsymbol{\theta}} = \min_{\boldsymbol{\theta}} \mathbb{E}_\Omega \left[ \mathbb{E}_{\widetilde{\Omega}} [\mathcal{L}(\{\widetilde{\mathbf{p}}'_s\}, \{\widetilde{\mathbf{p}}_t\})] \right] = \min_{\boldsymbol{\theta}} \mathbb{E}_\Omega \left[ \mathbb{E}_{\widetilde{\Omega}} \left[ \mathcal{L} \left( \left\{ \mathcal{T}_{\boldsymbol{\theta}}^{\{\mathbf{p}_s\} \rightarrow \{\mathbf{p}_t\}}(\widetilde{\mathbf{p}}_s) \right\}, \{\widetilde{\mathbf{p}}_t\} \right) \right] \right] \quad \text{(eq.5)}$$

where $\widetilde{\mathbf{p}}'_s = \mathcal{T}_{\boldsymbol{\theta}}^{\{\mathbf{p}_s\}\rightarrow\{\mathbf{p}_t\}}(\widetilde{\mathbf{p}}_s) = f_{\boldsymbol{\theta}_{trans}}\left(\widetilde{\mathbf{p}}_s \middle| f_{\boldsymbol{\theta}_{feat}}^{\{\mathbf{p}_s\}\rightarrow\{\mathbf{p}_t\}}\right)$ is parameterised by two neural networks, as described above; $\mathbb{E}_{\Omega}[\cdot]$ and $\mathbb{E}_{\widetilde{\Omega}}[\cdot]$ denote the expectation operators over the training point-set-pair domains $\Omega$ and $\widetilde{\Omega}$, training examples for $\{\mathbf{p}_s\}$ and $\{\mathbf{p}_t\}$ to compute the global feature, and training examples of $\{\widetilde{\mathbf{p}}_s\}$ and $\{\widetilde{\mathbf{p}}_t\}$ for computing the distance-based loss, respectively. This general form of training lets FPT provide the flexibility to allow the network input point-sets $\{\mathbf{p}_s\}$ and $\{\mathbf{p}_t\}$ to differ from the training "ground-truth" point-sets $\{\widetilde{\mathbf{p}}_s\}$ and $\{\widetilde{\mathbf{p}}_t\}$. This formulation can be applied in the following scenarios:

1. Unsupervised learning of point-set registration, i.e., $\{\widetilde{\mathbf{p}}_s\} = \{\mathbf{p}_s\}$ and $\{\widetilde{\mathbf{p}}_t\} = \{\mathbf{p}_t\}$.
2. Partial data registration with full data available in training, e.g., $\{\mathbf{p}_s\} \subseteq \{\widetilde{\mathbf{p}}_s\}$ or $\{\mathbf{p}_t\} \subseteq \{\widetilde{\mathbf{p}}_t\}$. This will provide a loss computed from, in general, stronger supervision $\{\widetilde{\mathbf{p}}_s\}$ and $\{\widetilde{\mathbf{p}}_t\}$, while test data at inference are more likely to have a different distribution that is similar to what is represented by $\{\mathbf{p}_s\}$ and $\{\mathbf{p}_t\}$.
3. Training-time bootstrap resampling (Saeed et al., 2020), when $\{\widetilde{\mathbf{p}}_s\}$, $\{\mathbf{p}_s\}$, $\{\widetilde{\mathbf{p}}_t\}$ or $\{\mathbf{p}_t\}$ is large or the difference between their sizes – i.e., the difference in the number of points – is large. This allows sampling a subset of any of these point-sets during a stochastic or mini-batch gradient descent while maintaining an unbiased gradient.
4. Weakly-supervised learning, i.e., $\{\mathbf{p}_s\} \supseteq \{\widetilde{\mathbf{p}}_s\}$ or $\{\mathbf{p}_t\} \supseteq \{\widetilde{\mathbf{p}}_t\}$, yet $\{\widetilde{\mathbf{p}}_s\}$ and $\{\widetilde{\mathbf{p}}_t\}$ are point sets with known point-to-point correspondence, which are available during training, while $\{\mathbf{p}_s\}$ and $\{\mathbf{p}_t\}$ are the network input available during inference.

### 2.2.1 Loss Functions

In this work, two loss functions are compared to measure the distance between two point-sets. The first, a widely used metric for determining the mean nearest neighbor distances between point-sets; the Chamfer Distance (Fan et al., 2017). Second, we employ the negative log-likelihood of a Gaussian mixture model (GMM) to encourage the network to minimize the difference between the distributions of the point-sets. A two-way Chamfer Distance is used in this work as follows:

$$\mathcal{L}_{CD}(\{\widetilde{\mathbf{p}}'_s\}, \{\widetilde{\mathbf{p}}_t\}) = \frac{1}{\widetilde{N}_t}\left(\sum_{t\in[1,\widetilde{N}_t]} \min_{s\in[1,\widetilde{N}_s]} \|\widetilde{\mathbf{p}}_t - \widetilde{\mathbf{p}}'_s\|_2^2\right) + \frac{1}{\widetilde{N}_s}\left(\sum_{s\in[1,\widetilde{N}_s]} \min_{t\in[1,\widetilde{N}_t]} \|\widetilde{\mathbf{p}}_t - \widetilde{\mathbf{p}}'_s\|_2^2\right) \text{ (eq.6)}$$

Unlike Chamfer Distance, the negative log-likelihood of a GMM is one of the alternatives which requires explicit parameters when considering outliers or noise levels. We assume that the spatially transformed source point-set $\{\widetilde{\mathbf{p}}'_s\}$ define the centres of the $\widetilde{N}_s$ Gaussian clusters in a mixture model with $\widetilde{N}_s + 1$ clusters, with the additional cluster being a uniform distribution (with a probability of $\frac{1}{\widetilde{N}_t}$) for potential outliers (Myronenko and Song, 2010). Given the target point-set $\{\widetilde{\mathbf{p}}_t\}$ as the model-fitting data, the GMM training loss can then be defined as the negative log-likelihood function of the mixture model, as follows:

$$\mathcal{L}_{GMM}(\{\widetilde{\mathbf{p}}'_s\}, \{\widetilde{\mathbf{p}}_t\}) = -\sum_{t=1}^{\widetilde{N}_t} \log\left(u\frac{1}{\widetilde{N}_t} + (1-u)\frac{1}{\widetilde{N}_s}\sum_{s=1}^{\widetilde{N}_s}\left(\frac{1}{(2\pi\sigma^2)^{\frac{D}{2}}}e^{-\frac{\|\widetilde{\mathbf{p}}_t - \widetilde{\mathbf{p}}'_s\|^2}{2\sigma^2}}\right)\right) \text{ (eq.7)}$$

where $\sigma^2$ is the isotropic covariance; $\frac{1}{\widetilde{N}_s}$ is the equal membership probabilities among the $\widetilde{N}_s$ Gaussian clusters that are defined by the source point-set; and $u, 0 \leq u \leq 1$, weights the uniform distribution. The loss function thus has two parameters, $\sigma^2$ and $u$.

As the FPT architecture does not explicitly constrain transformations from potential folding, collapse, or severe distortion of the transformed point-sets, the two-way construction of the loss functions, including both the Chamfer Distance and negative log-likelihood of a GMM, are employed in this work. It is interesting to find that, during the training, relaxing constraints such as 'one-to-one' correspondence did not cause unrealistic, extremely non-smooth deformation, based on point-set data extracted from real-world clinical images.

## 3 Experiments

### 3.1 Exemplar Clinical Application

Prostate MR-TRUS image fusion is a technique for using MR images to perform tumour-targeted needle biopsy (Moore et al., 2013; Marks et al., 2013) and minimally-invasive treatments (Dickinson et al., 2013) in patients for whom clinically significant prostate cancer is suspected or confirmed. The techniques involve presenting information on the location and size of MRI-visible lesions/tumours to complement the information provided by real-time TRUS images so that needles and other instruments can be placed to accurately target specific tissue regions. MRI-derived lesion/tumour information is typically displayed as a visual overlay superimposed on TRUS images, as a composite MR-TRUS image, or with the MR and TRUS images presented side-by-side. Displaying the images using any of these methods requires accurate registration. In this section, we used this registration task as an example to demonstrate how our FPT can be used for a real-world clinical application.

### 3.2 Data

The experimental dataset used in our evaluation comprised 108 pairs of pre-operative T2-weighted MR and intraoperative TRUS images from 76 patients which were acquired during the SmartTarget clinical trials (Hamid et al., 2019). Before point-set generation from the prostate contours, each of the MR and TRUS images was resampled to an isotropic voxel size of $0.8 \times 0.8 \times 0.8$ mm$^3$. Prostate gland boundaries were segmented in the resampled MR and TRUS images. Segmentations of the prostate gland in the MR images were acquired as part of the SmartTarget clinical trial protocols (Hamid et al., 2019). Additionally, segmentations of the prostate gland in the TRUS images were manually edited based on automatically contoured prostate glands from the original TRUS slices (Ghavami et al., 2018).

Using segmented MR and TRUS images, the contours and volumes of each prostate gland were extracted to generate two 3D point-sets – $P_T$ from the TRUS images and $P_S$ from the MR images – using a grid-based sampling approach in which each voxel was converted into a vector of its $[x, y, z]^\mathrm{T}$ 3D Euclidean coordinates in the segmented image volume. When the points are displayed, this gives the appearance of a grid-like point-set. As such, each voxel's location represented a single point in space in the generated point-set (Figure 2).

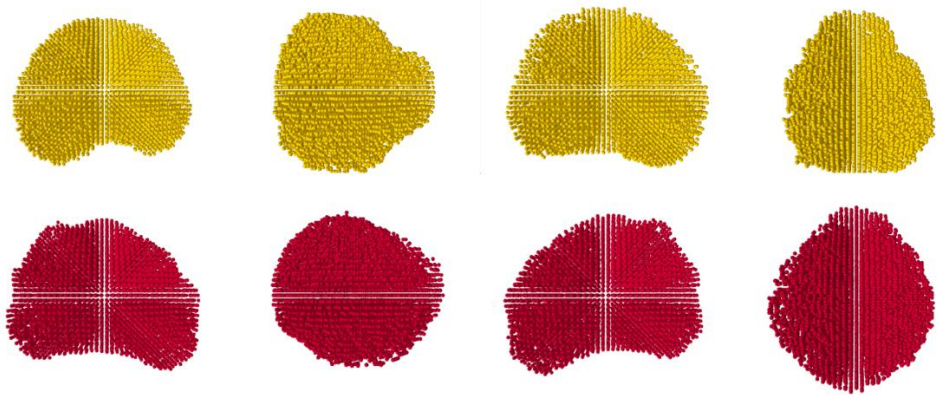

Figure 2: Point-sets depicting (from left to right) the anterior, right, posterior, and left views of a prostate volume obtained from a segmented TRUS (top) and MR volume (bottom) for one pair of patient images.

*3.3 Network Implementation and Training*

The previously described FPT was implemented in TensorFlow (Abadi et al., 2016) and Keras (Chollet, 2015). The FPT network architecture used in all experiments uses a value of $K = 1024$; thus, extracting a 1024-dimensional feature vector from each of the twin weight-sharing PointNets. In the point transformer module, $L = 6$, and $U = \{1024, 512, 256, 128, 64, 3\}$. The set of activation functions, $a$, was defined such that the first five layers used the ReLU activation function (Nair and Hinton, 2010), but the final layer used no activation function.

All networks were trained for 2000 epochs with the Adam optimizer (Kingma and Ba, 2015), a minibatch size of 8, and an initial learning rate of $1 \times 10^{-5}$ when training with the Chamfer Distance Loss ("FPT-Chamfer"). When training with the GMM Loss, all hyperparameters were identical to those used when training with the Chamfer Distance Loss, apart from a minibatch size of 2 ("FPT-GMM"). Additionally, hyperparameters $\sigma^2$ and $u$ in the GMM Loss were set as 0.001 and 0.1, respectively. Networks were trained on an NVIDIA DGX-1 system using a single Tesla V100 GPU.

During training, the points in $P_T$ and $P_S$ were permuted, scaled, and resampled on-the-fly. The points were scaled, per-sample, between $[-1, 1]$ in each of the $X$, $Y$ and $Z$ directions. Both point-sets were then shuffled and randomly subsampled to the desired cardinality. $P_S$ was further transformed on-the-fly with scaling, rotation, and displacement. Rotation angles were randomly sampled from $[-45°, 45°]$ about each of the $X$, $Y$ and $Z$ axes. Displacements were randomly sampled from $[-1, 1]$ in each of the $X$, $Y$ and $Z$ directions.

*3.4 Gaussian Radial Basis Function Network*

To demonstrate the effectiveness of FPT, we compare to the use of a parametric transformation, similar to that proposed by Fu et al. (2021). This network uses a Gaussian Radial Basis Function (G-RBF) model to compute, and account for, the complex deformation between the surfaces of the source and target point-sets. In our implementation of the G-RBF network, the global feature extraction module developed for FPT is used, however; we replace our point transformation module with a G-RBF module. This G-RBF module determines the point displacements by predicting the control points and spline coefficients. Subsequently, the input point-sets are registered by non-rigid transformation via computation of the displacement between source and target point clouds, providing the transformed source points as

$$\widetilde{\mathbf{p}}'_s = f_{\theta_{GRBF}}(\widetilde{\mathbf{p}}_s) = \boldsymbol{\alpha} \mathbf{k}_s + \widetilde{\mathbf{p}}_s \text{ (eq. 8)}$$

where $\mathbf{k}_s = [k_s^1, k_s^2, \ldots, k_s^{N_c}]^T$ is the $N_c \times 1$ Gaussian kernel vector $k_s^c(\mathbf{p}_c, \widetilde{\mathbf{p}}_s) = \exp\left(-\frac{\|\mathbf{p}_c - \widetilde{\mathbf{p}}_s\|^2}{2\sigma^2}\right)$ (Fornefett et al., 2001), computed for each point $\widetilde{\mathbf{p}}_s$, with respect to a set of control points $\mathbf{p}_c$, $c = 1, 2, \ldots, N_c$. The $N_c$ control points $\mathbf{p}_c$ and the $3 \times N_c$ spline coefficients $\boldsymbol{\alpha}$ are directly predicted by the G-RBF point transformation network. In this work, the G-RBF uses $N_c = 27$ control points and a kernel parameter $\sigma = 1$, unless otherwise indicated. The G-RBF network was trained with the same amount of training data, the same data augmentation methods, and the same loss functions as the FPT.

Two variants of the proposed G-RBF network were compared, each with different loss functions used in training: First, using the Chamfer Distance Loss ("G-RBF-Chamfer"), and secondly, using the ("G-RBF-GMM"), both as previously described in Section 2.2.1. An illustration of the G-RBF network, including the G-RBF point transformation module is presented in Figure 3.

*3.5 Comparison with G-RBF and Classical Methods*

In addition to the previously described G-RBF networks, the FPT was compared to two example pairwise iterative methods for point-set registration: the widely used rigid ICP algorithm and the non-rigid CPD algorithm. In our experiments, the ICP algorithm was permitted to complete up to 25 iterations. All other parameters and initializations were performed as described by Besl and McKay (Besl and McKay, 1992). For the CPD algorithm, we set $w$ = 0, where the value of $w$ ($0 \leq w \leq 1$) indicates the assumed amount of noise present in the point-set and permitted the algorithms to complete up to 250 iterations. All other parameters were set to the default values described by Myronenko and Song (Myronenko and Song, 2010).

A series of experiments were performed to demonstrate the FPT's performance compared to the G-RBF networks (G-RBF-Chamfer and G-RBF-GMM), ICP, and CPD for registration of MR to TRUS data. In these experiments, we utilized the dataset described above, which comprises 108 pairs of MR and TRUS images. These data were split into a training and testing set, wherein 70% of the data (75 MR-TRUS pairs) were reserved for training, and 30% of the data (33 MR-TRUS pairs) were reserved for testing. Any patient who had multiple series of imaging was assigned to the training set to ensure that images from a single patient were not included in both the training and the test set. We did not use a hold-out set to prevent bias by an exhaustive hyperparameter search when creating and training FPT to demonstrate the ability of its data-driven architecture compared to other methods. It should be noted that this two-way split experiment may systematically underestimate the registration performance of the G-RBF network and other methods which rely on extensive hyper-parameter tuning.

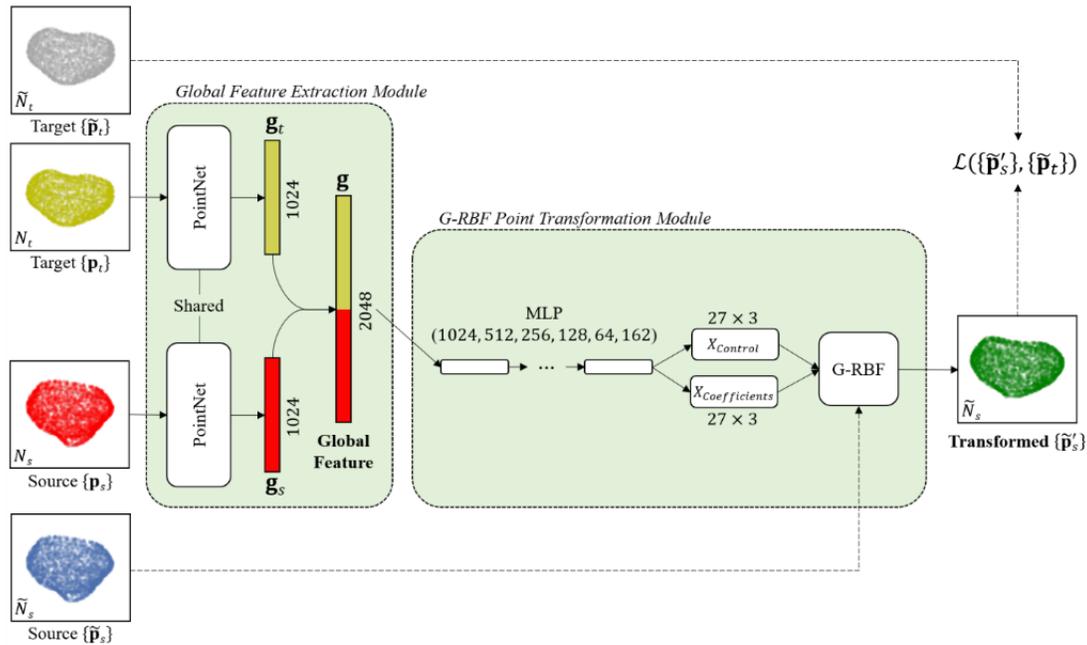

Figure 3: Schematic representation of the G-RBF network design for non-rigid point-set registration. Similarly to FPT, the global feature extraction module takes a target and source point-set and applies shared input and feature transformations to both, creating a global feature vector. The G-RBF point transformation module applies a non-rigid transformation using the predicted control points and spline coefficients, as in Eq. 8, to obtain the transformed point-set. MLP stands for multi-layer perceptron. G-RBF stands for Gaussian Radial Basis Function.

### 3.5.1 MR to TRUS Registration

We evaluated FPT's non-rigid registration performance when aligning complete volumetric MR and TRUS point-sets, similar to the first scenario presented in Section 2.2. Performance in this registration problem was tested by varying the size of $\{\mathbf{p}_s\}$ and $\{\widetilde{\mathbf{p}}_s\}$. Both the FPT and our G-RBF network were trained using both loss functions, using 1024, 2048, 4096, or 8192 points in $\{\mathbf{p}_s\}$. Owing to the cardinality invariance of the FPT and G-RBF network architectures, each of these trained networks was then used to predict registrations with 1024, 2048, 4096, or

8192 points in $\{\widetilde{\mathbf{p}}_s\}$ to test the sensitivity to different sampling rates between network inputs during training and at inference. The ICP and CPD algorithms do not require training and were evaluated on the computed registrations they produced with 1024, 2048, 4096, or 8192 input points.

*3.5.2 MR to Partial TRUS Registration*

Additionally, we evaluated FPT's non-rigid registration performance when aligning complete volumetric MR point-sets to partial volumetric TRUS point-sets, similar to the second scenario presented in Section 2.2. This experiment was designed to reflect three clinical scenarios in which only point-data defining part of the prostate surface are available due to a small number of 2D TRUS images being acquired, each representing a different slice through the organ. For each of these scenarios, surface points from only two or three segmented ultrasound slices were used as inputs to the registration algorithms.

Examples of each prostate TRUS imaging scenario investigated are illustrated in Figure 4. The first scenario represents the case where three evenly distributed 2D TRUS slices are obtained. The second scenario represents the case where two or three TRUS slices are obtained, but the slices are biased to one lateral direction. The third scenario represents the case where only two TRUS slices are obtained which provide poor coverage of the prostate gland, with the slices skewed to the left or right side.

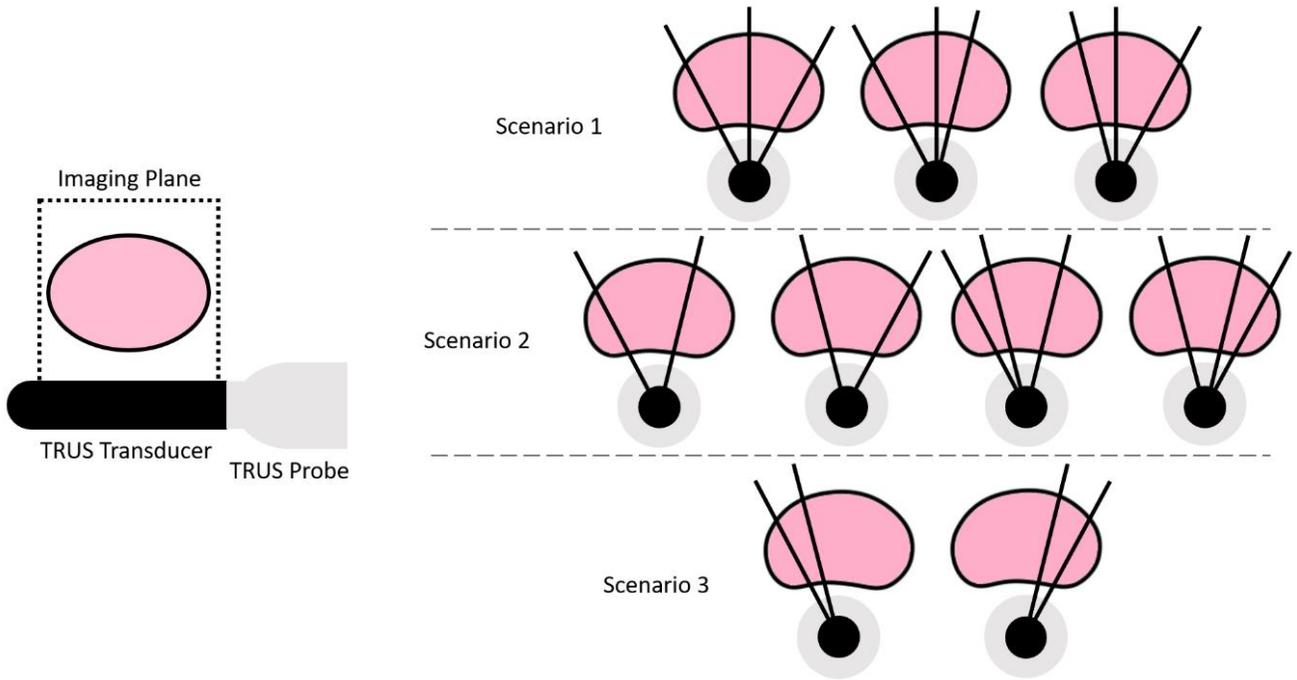

Figure 4: Illustration of possible TRUS images acquired from the TRUS transducer. Images are captured in the sagittal plane (left) and are shown with other image slices that may be acquired in each of the three scenarios, from an axial view (right).

To quantitatively describe and validate the differences between each scenario, we used two metrics: slice centroid distance and slice span. Slice centroid distance was defined as follows:

$$Slice\ Centroid\ Distance = \|\mathbf{c}_p - \mathbf{c}_s\|_2 \quad (eq.\ 9)$$

where $\mathbf{c}_p$ is the geometric center of the TRUS prostate point-set, and $\mathbf{c}_s$ is the geometric center of the point-set comprising all the selected TRUS image slices. Additionally, slice span was defined as follows:

$$Slice\ Span = \sqrt{\frac{1}{n}\sum_{i=1}^{n}\|\mathbf{c}_p - \mathbf{c}_s^i\|_2^2} \quad (eq.\ 10)$$

where the set $\{\mathbf{c}_s^i \mid i = 1, \dots, n\}$ describes the centroid points comprising the individual selected TRUS image slices from $n$ slices. These metrics are illustrated in Figure 5 and expected and computed values for the metrics which quantitatively describe the distribution of points and individual frames in each scenario are given in Table 1.

Table 1: Qualitative metrics which describe the three clinical scenarios. Values: Mean ± Std.

|  | Slice Centroid Distance (mm) |  | Slice Span (mm) |  |
| --- | --- | --- | --- | --- |
|  | Expected | Actual | Expected | Actual |
| Scenario 1 | Lowest | 4.07 ± 0.94 | Highest | 10.40 ± 1.28 |
| Scenario 2 | Between Scn. 1 & Scn. 3 | 4.78 ± 1.43 | Between Scn. 1 & Scn. 3 | 8.45 ± 1.31 |
| Scenario 3 | Highest | 10.6 ± 2.63 | Lowest | 4.74 ± 1.05 |

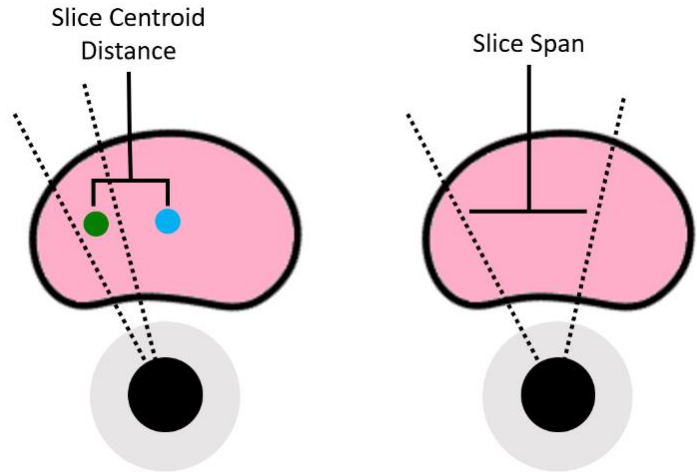

Figure 5: Illustration of the Slice Centroid Distance (left) and Slice Span (right) in two simulated instances of the possible scenarios. The green point indicates the centroid of the points in the selected image slices, and the blue point indicates the centroid of the entire prostate volume.

In the first set of experiments, we observe that changing the point sampling rates in training and at inference does not affect our selected registration metrics (see Section 3.5). Therefore, in this second set of experiments, we only train instances of FPT-Chamfer and G-RBF-Chamfer with input point-sets containing 4096 points in each of the three TRUS scenarios. An input size of 4096 points was selected empirically as the previous experiment demonstrated no clear difference in registration quality when varying input point-set sizes. The Chamfer Distance Loss was selected as it reduced training time and produced superior results for registration error when compared to FPT networks trained with the GMM Loss in the previous experiment. For additional comparison, the ICP and CPD algorithms were also evaluated in these three previously described scenarios with input point-sets containing 4096 points.

### 3.6 Evaluation of Registration Methods

The accuracy of the prostate surface point registrations was quantified using the Chamfer distance, the Hausdorff distance, and a target registration error (TRE), calculated as the distance between points defining the 3D locations of corresponding, manually identified anatomical landmarks in the TRUS and MR images (Hu et al., 2018; Ghavami et al., 2019). The Chamfer distance was used as the loss function for some of the experiments, and therefore indicates the network generalisation to independent test data. Together with the Hausdorff distance, the registration accuracy on the point-set-represented individual prostate glands can be measured. The TRE is defined as the root-mean-square of each of the distances computed between the geometric centroids of the registered pairs of source and target landmarks for each patient. The landmarks in our dataset comprised 309 pairs of points, including points defining the apex and base of the prostate, and various patient-specific landmarks including zonal structure boundaries, water-filled cysts, and calcifications (Hamid et al., 2019). It should be noted that the overall spatial distribution of these landmarks may be representative of the full TRE distribution in this application (Hu et al., 2008; Hahn et al., 2010; Karnik et al., 2010; Hu et al., 2011; Heinrich et al., 2012; Hu et al., 2012; Mitra et al, 2012a; Mitra et al., 2012b; De Silva et al., 2013; Rivas et al., 2014; Sun et al., 2014; Fedorov et al., 2015; Hu et al., 2015; Sun et al., 2015; Yang et al., 2015; Zettinig et al., 2015; Wang et al., 2016 Onofrey et al., 2017; Hu et al., 2018; Sun et al., 2018; Yan et al., 2018; Sultana et al., 2019; Xu et al., 2020; Fu et al., 2021; Hu et al., 2021), but landmark-based TREs nevertheless provide a useful estimate of the errors associated with localising tumours within the prostate. The computational time was also recorded for each registration experiment.

## 4 Results

### 4.1 MR to TRUS Registration

Table 2 shows the mean and standard deviation for Chamfer distance, Hausdorff distance and TRE for each of the different methods and input point-set sizes. Across all variants and experiments in MR to TRUS registration, FPT-Chamfer achieves a mean TRE of 4.71 mm, compared to 5.16 mm for FPT-GMM, 5.29 mm for G-RBF-Chamfer, 5.25 mm for G-RBF-GMM, 6.02 mm for ICP, and 5.08 mm for CPD. Without any form of alignment on the dataset, we observe a TRE of 25.43 mm. FPT-Chamfer gives the lowest average Chamfer distance and TRE in nearly all instances, while CPD gives the lowest average for Hausdorff distance. The prostate contours from a sample slice in the transverse plane from resulting registrations of three cases with each of the learning-based and iterative methods are shown in Figure 6.

Table 2: Chamfer distance, Hausdorff distance, and TRE for each method used and at each point-set size in the first MR-TRUS registration experiment. Values: Mean ± Std. The lowest mean value in each section is bolded. Significant differences with respect to FPT-Chamfer are denoted with an asterisk (*), based on two-tailed paired t-tests at α = 0.05.

| Number of Points in $\{\widetilde{\mathbf{p}}_s\}$ | Method | Chamfer Distance (mm) | Hausdorff Distance (mm) | TRE (mm) |
| --- | --- | --- | --- | --- |
| 8192 | FPT-Chamfer | **1.10 ± 0.17** | 6.18 ± 1.36 | **4.75 ± 1.45** |
|  | FPT-GMM | 1.14 ± 0.18 | 6.65 ± 1.53* | 5.49 ± 1.68* |
|  | G-RBF-Chamfer | 1.15 ± 0.17 | 7.20 ± 1.43* | 5.50 ± 1.61* |
|  | G-RBF-GMM | 1.17 ± 0.16* | 7.70 ± 2.00* | 4.87 ± 1.33 |
|  | ICP | 1.29 ± 0.19* | 8.48 ± 1.79* | 5.94 ± 1.68* |
|  | CPD | 1.25 ± 0.19* | **6.12 ± 1.32** | 5.12 ± 1.35 |
| 4096 | FPT-Chamfer | **1.38 ± 0.19** | 6.27 ± 1.41 | **4.69 ± 1.41** |
|  | FPT-GMM | 1.41 ± 0.20 | 6.66 ± 1.36* | 5.34 ± 1.50* |
|  | G-RBF-Chamfer | 1.42 ± 0.19 | 7.30 ± 1.48* | 4.92 ± 1.34 |
|  | G-RBF-GMM | 1.45 ± 0.21* | 7.83 ± 1.66* | 5.19 ± 1.46* |
|  | ICP | 1.53 ± 0.22* | 8.58 ± 1.86* | 6.03 ± 1.62* |
|  | CPD | 1.44 ± 0.20 | **5.98 ± 1.34** | 4.99 ± 1.39 |
| 2048 | FPT-Chamfer | **1.72 ± 0.21** | 6.46 ± 1.20 | **4.55 ± 1.34** |
|  | FPT-GMM | 1.75 ± 0.23 | 6.81 ± 1.32 | 4.70 ± 1.52 |
|  | G-RBF-Chamfer | 1.77 ± 0.23* | 7.50 ± 1.56* | 5.42 ± 1.75* |
|  | G-RBF-GMM | 1.80 ± 0.24* | 7.94 ± 1.78* | 5.52 ± 1.40* |
|  | ICP | 1.89 ± 0.23* | 8.63 ± 1.84* | 6.09 ± 1.53* |

|      | Method       |                 |                |                |
|------|--------------|-----------------|----------------|----------------|
|      | CPD          | 1.73 ± 0.22     | **6.07 ± 1.13*** | 4.98 ± 1.42*   |
| 1024 | FPT-Chamfer  | 2.16 ± 0.29     | 6.73 ± 1.22    | **4.83 ± 1.42** |
|      | FPT-GMM      | 2.19 ± 0.29*    | 7.10 ± 1.30    | 5.09 ± 1.40    |
|      | G-RBF-Chamfer| 2.20 ± 0.30*    | 7.82 ± 1.56*   | 5.34 ± 1.64*   |
|      | G-RBF-GMM    | 2.24 ± 0.30*    | 8.16 ± 1.79*   | 5.40 ± 1.89*   |
|      | ICP          | 2.34 ± 0.33*    | 9.17 ± 2.03*   | 6.01 ± 1.79*   |
|      | CPD          | **2.10 ± 0.25*** | 6.49 ± 1.42   | 5.21 ± 1.34*   |

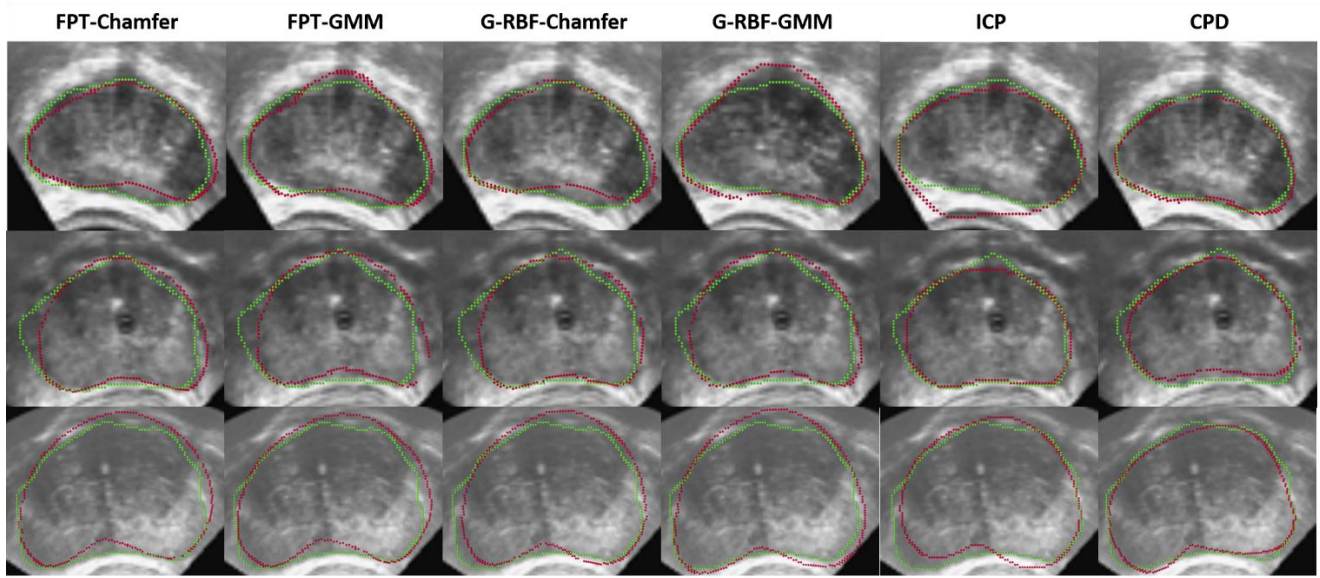

Figure 6: Example image slices illustrating registration results from three different cases, shown in the transverse plane. Each image shows the original TRUS image with the transformed source (MR) contours (red) superimposed on to the target (TRUS) contours (green). Columns indicate registrations from each method; from left to right: FPT-Chamfer, FPT-GMM, G-RBF-Chamfer, G-RBF-GMM, ICP, and CPD.

For our FPT-Chamfer and FPT-GMM implementations, between 14 and 50 registrations may be performed per second, depending on the size of the input point-set at inference. These registration times are nearly identical to those achieved with the G-RBF network (G-RBF-Chamfer and G-RBF-GMM), approximately 5-8 times faster than those observed with ICP, and approximately 200-5000 times faster than those observed with CPD. Table 3 shows the mean and standard deviation of the registration times for the different methods.

Table 3: Time to compute a single registration at a given point-set size for FPT, G-RBF, ICP and CPD. Values: Mean ± Std.

|        | Number of Points in $\{\widetilde{\mathbf{p}}_s\}$ | | | |
|--------|------|------|------|------|
| Method | 1024 | 2048 | 4096 | 8192 |
| FPT    | 0.02s ± 0.00 | 0.02s ± 0.01 | 0.04s ± 0.01 | 0.07s ± 0.01 |
| G-RBF  | 0.02s ± 0.00 | 0.02s ± 0.01 | 0.04s ± 0.01 | 0.07s ± 0.01 |
| ICP    | 0.10s ± 0.01 | 0.15s ± 0.02 | 0.31s ± 0.02 | 0.45s ± 0.02 |
| CPD    | 4.17s ± 1.04 | 17.89s ± 3.15 | 85.58s ± 10.52 | 357.73s ± 25.74 |

To assess if the changes in Chamfer Distance and Hausdorff Distance when using different numbers of points at inference is related to the size and inherent point density of $\{\widetilde{\mathbf{p}}_s\}$, we also report the results of MR to TRUS registrations performed on a grouped series of random subsamples without replacement. By creating multiple unique and non-intersecting subsets of points for each MR and TRUS prostate volume, we can group each of the predicted registrations from these subsets. This was performed at four different thresholds for the size of $\{\widetilde{\mathbf{p}}_s\}$; we performed eight registrations of eight subsets of 1024 points, four registrations on four subsets of 2048 points, two registrations on two subsets of 4096 points, and one registration on the original 8192 points.

We summarize the results of these registrations by presenting the mean and standard deviation of Chamfer distance, Hausdorff distance, and TRE at the four different thresholds in Table 4. We observe that grouping registrations with different point sampling rates at inference does not appear to affect Chamfer distance or Hausdorff distance. This demonstrates that differences in Chamfer Distance and Hausdorff Distance in our prior experiments may be due to point-set density; where a less dense point-set produces a higher value for the same metrics. We conclude that, with sufficient points sampled in each set, the obtained TRE became less sensitive to the tested different sampling strategies and increase of the sampling density, a practically desirable property of the proposed method. While there are small variations in the average reported TRE for the grouped registrations, FPT-Chamfer still produces the lowest overall average TRE in the registrations performed at each threshold for the size of $\{\widetilde{\mathbf{p}}_s\}$.

Table 4: Chamfer distance, Hausdorff distance, and TRE for each method used and at each grouped registration threshold. Values: Mean ± Std.

| Number of Points in $\{\widetilde{\mathbf{p}}_s\}$ | Grouped Registrations | Method | Chamfer Distance (mm) | Hausdorff Distance (mm) | TRE (mm) |
|---|---|---|---|---|---|
| 8192 | 1 | FPT-Chamfer | 1.10 ± 0.17 | 6.03 ± 1.35 | 4.80 ± 1.28 |
| | | FPT-GMM | 1.16 ± 0.18 | 6.82 ± 1.49 | 5.33 ± 1.70 |
| | | G-RBF-Chamfer | 1.15 ± 0.18 | 7.33 ± 1.43 | 5.43 ± 1.73 |
| | | G-RBF-GMM | 1.18 ± 0.17 | 7.88 ± 1.99 | 5.58 ± 1.40 |
| 4096 | 2 | FPT-Chamfer | 1.10 ± 0.17 | 6.20 ± 1.34 | 4.74 ± 1.25 |
| | | FPT-GMM | 1.12 ± 0.16 | 6.60 ± 1.47 | 5.37 ± 1.68 |
| | | G-RBF-Chamfer | 1.14 ± 0.16 | 7.21 ± 1.45 | 5.21 ± 1.76 |
| | | G-RBF-GMM | 1.17 ± 0.19 | 7.76 ± 2.04 | 5.12 ± 1.58 |
| 2048 | 4 | FPT-Chamfer | 1.10 ± 0.16 | 6.37 ± 1.36 | 4.67 ± 1.22 |
| | | FPT-GMM | 1.12 ± 0.17 | 6.30 ± 1.48 | 4.94 ± 1.59 |
| | | G-RBF-Chamfer | 1.13 ± 0.19 | 7.07 ± 1.46 | 5.28 ± 1.69 |
| | | G-RBF-GMM | 1.16 ± 0.18 | 7.57 ± 2.01 | 5.35 ± 1.68 |
| 1024 | 8 | FPT-Chamfer | 1.09 ± 0.17 | 6.00 ± 1.33 | 4.79 ± 1.27 |
| | | FPT-GMM | 1.11 ± 0.18 | 6.31 ± 1.50 | 5.23 ± 1.57 |
| | | G-RBF-Chamfer | 1.15 ± 0.19 | 7.23 ± 1.39 | 5.18 ± 1.62 |
| | | G-RBF-GMM | 1.16 ± 0.18 | 7.55 ± 1.96 | 5.21 ± 1.56 |

### 4.2  MR to Partial TRUS Registration

Table 5 shows the mean and standard deviation for Chamfer distance, Hausdorff distance and TRE for each of the different methods in each different clinical scenario. Across all methods and scenarios in the MR to partial TRUS registration, FPT-Chamfer achieves the lowest average Chamfer distance, Hausdorff distance and TRE in all instances. Among the deep learning-based methods, average Chamfer distance, Hausdorff distance, and TRE are similar to those in the first series of experiments where complete data were available and $\{\widetilde{\mathbf{p}}_s\}$ contained 4096 points. ICP and CPD demonstrate lower average performance in all metrics relative to their results in the previous experiment. Most saliently, Chamfer distance and Hausdorff distance for CPD are 1.5-6 times higher on average than in the previous experiment, and computed values of TRE nearly double on average. Based on two-tailed paired t-tests at α = 0.05, the differences in Hausdorff Distance and TRE across all three scenarios between FPT-

Chamfer and G-RBF-Chamfer, FPT-Chamfer and ICP, and FPT-Chamfer and CPD are statistically significant ($p < 0.005$, $p < 0.005$, and $p < 0.001$, respectively). The differences in Chamfer Distance across all three scenarios between FPT-Chamfer and ICP, and FPT-Chamfer and CPD are also statistically significant ($p < 0.005$, and $p < 0.001$, respectively). 3D visualizations of the prostate shapes before and after registration with variants of FPT for three different cases are given in Figure 7. The prostate contours from a sample slice in the transverse plane from resulting registrations of three cases with each of the scenarios for FPT-Chamfer are shown in Figure 8. A box plot of the TREs at comparing FPT-Chamfer, G-RBF-Chamfer, ICP, and CPD at the patient level for MR to TRUS and MR to partial TRUS registrations in all three scenarios is given in Figure 9.

Table 5: Chamfer distance, Hausdorff distance, and TRE for each method used in the partial MR-TRUS registration experiment. All point-sets are of size 4096. Values: Mean ± Std. The lowest mean value in each section is bolded. Significant differences with respect to FPT-Chamfer are denoted with an asterisk (*), based on two-tailed paired t-tests at α = 0.05.

| Scenario | Method | Chamfer Distance (mm) | Hausdorff Distance (mm) | TRE (mm) |
|---|---|---|---|---|
| Scenario 1 | FPT-Chamfer | **1.40 ± 0.20** | **6.38 ± 1.48** | **4.88 ± 1.56** |
| | G-RBF-Chamfer | 1.45 ± 0.20 | 7.38 ± 1.68* | 5.39 ± 1.79* |
| | ICP | 1.93 ± 0.41* | 9.07 ± 1.95* | 6.54 ± 2.19* |
| | CPD | 2.25 ± 0.42* | 15.74 ± 4.29* | 9.35 ± 3.04* |
| Scenario 2 | FPT-Chamfer | **1.41 ± 0.21** | **6.36 ± 1.70** | **4.81 ± 1.75** |
| | G-RBF-Chamfer | 1.46 ± 0.22 | 7.68 ± 1.73* | 5.27 ± 1.95* |
| | ICP | 1.94 ± 0.43* | 9.28 ± 2.24* | 6.48 ± 2.24* |
| | CPD | 3.32 ± 0.72* | 21.38 ± 6.28* | 9.69 ± 3.36* |
| Scenario 3 | FPT-Chamfer | **1.42 ± 0.21** | **6.62 ± 1.90** | **4.76 ± 1.71** |
| | G-RBF-Chamfer | 1.45 ± 0.23 | 7.71 ± 2.11* | 5.55 ± 2.38* |
| | ICP | 1.94 ± 0.62* | 9.12 ± 2.59* | 7.04 ± 2.33* |
| | CPD | 6.58 ± 1.03* | 36.78 ± 7.27* | 10.30 ± 3.74* |

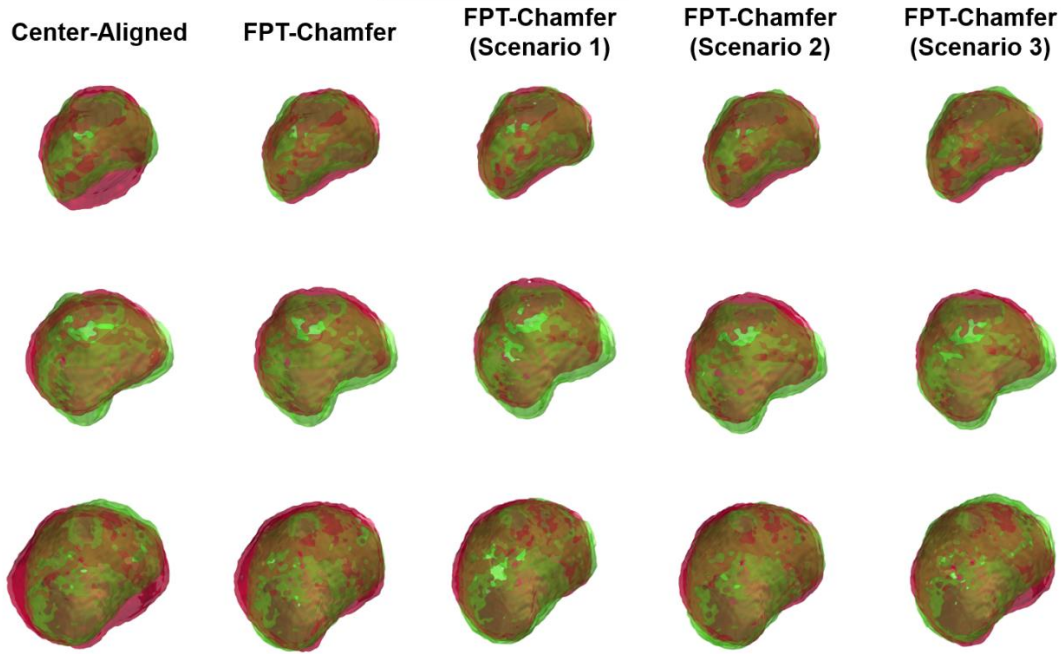

Figure 7: MR-TRUS prostate glands showing the overlap of the transformed source (MR) and target (TRUS) point-sets. The green shape illustrates the target point-set, while the red shape illustrates the transformed source point-set. The first column shows the source and target after center-alignment. The remaining columns show registrations from various methods; from left to right: FPT-Chamfer, FPT-Chamfer trained with Scenario 1 image slices, FPT-Chamfer trained with Scenario 2 image slices, and FPT-Chamfer trained with Scenario 3 image slices.

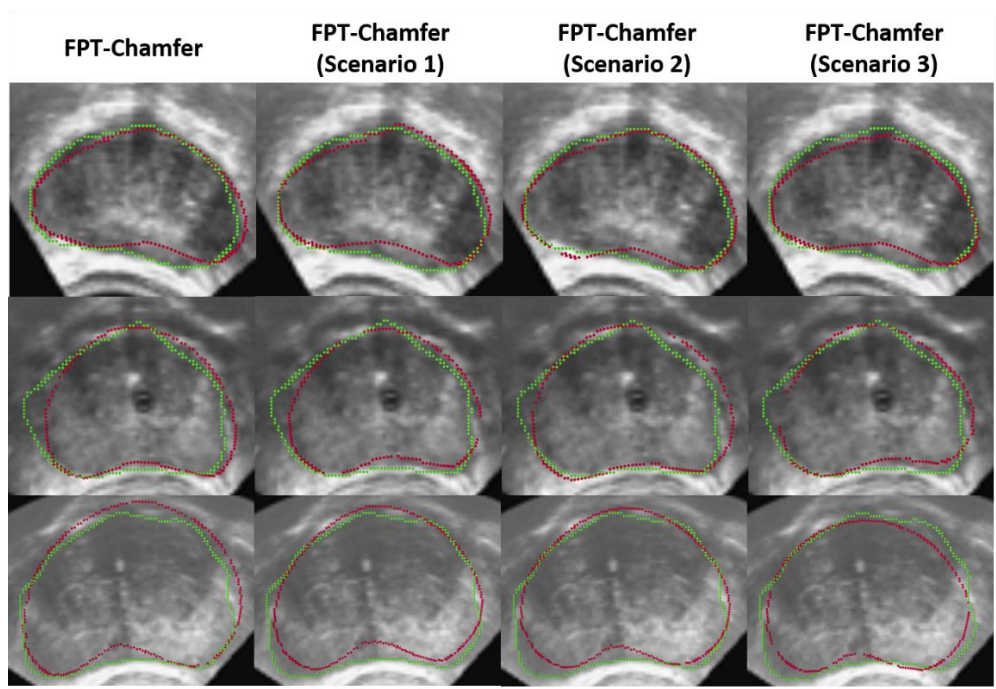

Figure 8: Example image slices illustrating registration results from three different cases, shown in the transverse plane. Each image shows the original TRUS image with the transformed source (MR) contours (red) superimposed on to the target (TRUS) contours (green). Columns indicate registrations from FPT-Chamfer for registrations with; from left to right; full volumes, Scenario 1, Scenario 2, and Scenario 3.

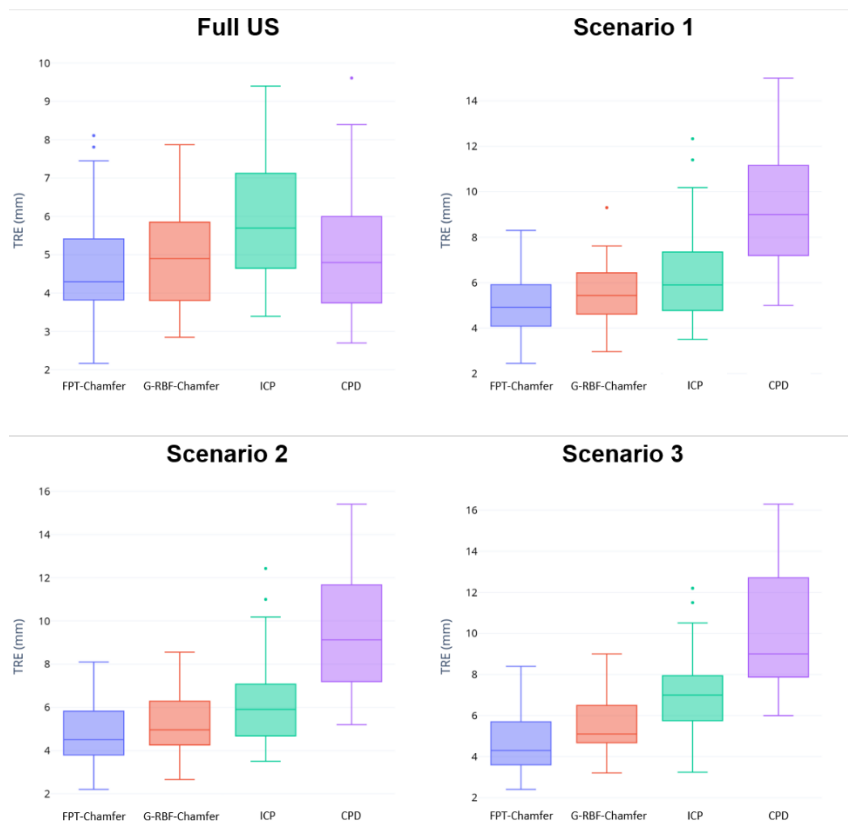

Figure 9: Boxplots of the root-mean-square TRE per patient for the MR to TRUS and MR to partial TRUS registrations obtained from FPT-Chamfer, G-RBF-Chamfer, ICP, and CPD with 4096 points.

## 5 Discussion

In this work, we proposed a deep learning-based approach for point-set registration, called FPT, and apply it to an example multimodal image registration problem, namely that of prostate MR-TRUS registration. Unlike intensity-based methods, wherein similarity metrics are often utilized, the FPT leverages the geometric and spatial information from point-sets to drive the learning, and subsequent registration process. While the effectiveness of our work relies on the extraction of features from the imaging data, the point-sets required may be generated efficiently and automatically via accurate image segmentations obtained from emerging deep learning methods (Ghavami et al., 2018; Fu et al., 2021). For TRUS-MRI registration, only a few 2D US images may need to be segmented to produce a sufficient number of input points for registration using the aforementioned grid-based sampling approach described in this work. Furthermore, using FPT-Chamfer, TREs are lower or comparable to all other methods, with a mean TRE in the first and second experiments of 4.71 mm and 4.81 mm, respectively. As illustrated in Table 5, FPT-Chamfer significantly outperforms other methods in the partial registration, in all metrics, except for G-RBF-Chamfer, with a two-way Chamfer Distance loss (eq. 6), when evaluating also using Chamfer Distance. Other independent metrics, including Hausdorff Distance and TRE, have all supported the superior generalisation ability from FPT, with statistical significance.

We refer to our previous work for a comprehensive report of the quantitative results using full data sets, i.e., full 3D segmentation of the prostate in training with complete volumetric TRUS volume available at inference (Hu et al., 2018; Ghavami et al., 2019). In these works, the reported TREs were in general lower, between 3.0 and 3.6 mm, with a significant variance that led to a TRE of 6.1 mm with one of the reported G-RBF networks (Hu et al., 2018). As such, the displacements learned from point-set features in prostate gland segmentations may reduce variance in registration error, although additional validation is needed to draw further conclusions.

Though there is a measurable difference in the mean TRE, Chamfer distance, and Hausdorff distances obtained between the full volumetric registrations and partial registrations for FPT-Chamfer, it is notable that these variations provide little qualitative difference, as seen in Figures 7 and 8; only sub-millimetre differences in quantitative performance were observed between each of the three clinical scenarios. This highly comparable performance demonstrates FPT's flexibility and generalizability between different input data and illustrates that the network can adapt to multiple varied distributions and availabilities of input data and still learn to predict a desirable registration.

Intensity-based methods for multimodal image registration are also able to utilize information from the entire prostate gland, typically providing a qualitatively and quantitatively good intraprostatic deformation. To emulate this, we utilize volumetric point-sets which allows the network to learn intraprostatic deformation instead of relying on surface-driven deformations to interpolate intraprostatic deformation, which may result in unlikely interior deformations.

Recently, intensity-based deep learning methods have achieved reported TREs below 5 mm for the MRI-TRUS registration application explored in this paper (Hu et al., 2018; Sun et al., 2018; Yan et al., 2018; Sultana et al., 2019;). The TREs for this application estimated in this work fall within the expected range found in the literature, however, it is difficult to make direct comparisons between our results and other methods due to variations in the quality of data (for example, due to different clinical setups, image acquisition protocols, and user experience) and validation methods. In particular, the number and spatial distribution of landmarks used to estimate a TRE is likely to have a significant impact on the numerical error. In our dataset, the landmarks used to calculate TRE, such as are the apex and base of the prostate, are located on the surface or towards the periphery of the prostate gland (unlike the urethra, for instance). Furthermore, the aforementioned works do not consider the practical scenario of primary interest in this work, where only partial data are available due to a limited number of image slices. An important finding of this study is the minimal impact of using partial point data on accuracy when using the FPT method compared to other methods tested. Without extensive validation, It is unclear if the performance of intensity-based methods and/or other forms of representations, such as binary masks, would also be minimally impacted by this reduction of data. As such, the practical effects of partial data when applied to existing registration methods and frameworks merit a thorough validation and assessment but are considered out the scope of this work given the inherent challenges associated with successfully modifying these methods to represent and accurately register partial data. These results demonstrate that the FPT can learn descriptive, data-driven features directly from partial data without compromising registration accuracy. Unlike conventional image-based registration methods, these features enable efficient computation of a set of accurate displacements without cost-prohibitive hardware and rapidly enough to be suitable for real-time applications.

An important direction to further this work is to test the feature-based methods' ability to develop modality-, protocol-, and scanner-independent registration methods, owing to their non-reliance on the direct imaging data; a wider patient population, coming from multiple centers and with different acquisition protocols wherein there may be increased data heterogeneity is of value in future validation of our presented methods.

Given its robust, generalizable, and rapid point-set registration approach, there is potential that FPT will be of use in various other applications. Beyond MR-TRUS registration, FPT may be of use as an alternative to classical methods for general feature-based, non-rigid registration applications where point-sets may be reliably extracted.

## 6  Conclusion

We have introduced FPT, a novel approach to point-set registration using deep neural networks which learns the displacement field required to produce individual point displacements. Through evaluation in a challenging real-world multimodal image registration task with MR and TRUS images, FPT was found to be robust to the partial availability of data. Furthermore, this work demonstrates that partially available data, generated from automatically segmented MR and TRUS images, may be used to enable continual real-time MR-TRUS image registration during prostate biopsies. Such an architecture is of interest in other medical imaging problems where training data is limited or only partially available, as FPT may permit rapid registration of point-sets extracted from other types of imaging modalities as well. As a method for point-set registration which non-iteratively performs non-rigid registration without the need to establish point correspondence, we believe that FPT represents significant progress towards a generally applicable method for learning-based non-rigid registration in medical imaging.

## Acknowledgements

Z.M.C. Baum is supported by the Natural Sciences and Engineering Research Council of Canada Postgraduate Scholarships-Doctoral Program, and the University College London Overseas and Graduate Research Scholarships. This research was funded in whole, or in part, by the Wellcome Trust [203145Z/16/Z]. For the purpose of Open Access, the author has applied a CC BY public copyright licence to any Author Accepted Manuscript version arising from this submission.

## Declaration of Competing Interest

The authors declare that they have no known competing financial interests or personal relationships that could have appeared to influence the work reported in this paper.

## CRediT authorship contribution statement

Zachary M C Baum: Conceptualization, Methodology, Software, Validation, Formal Analysis, Visualization, Writing - Original Draft, Writing – Review & Editing. Yipeng Hu: Conceptualization, Methodology, Data Curation, Writing - Original Draft, Writing – Review & Editing, Supervision. Dean C Barratt: Supervision, Writing – Review & Editing.